\RequirePackage[2020-02-02]{latexrelease}
\documentclass[aps,11pt,preprintnumbers,nofootinbib,floatfix]{revtex4}

\usepackage{subfigure}
\usepackage{multirow}
\usepackage{amsmath}

\usepackage{amssymb}
\usepackage{epsfig}
\usepackage{bbold}
\usepackage{graphicx}

\begin{document}
\title{Prediction of non-SUSY AdS conjecture on the lightest neutrino mass revisited}
\author{Cao H. Nam}
\email{nam.caohoang@phenikaa-uni.edu.vn}  
\affiliation{Phenikaa Institute for Advanced Study and Faculty of Fundamental Sciences, Phenikaa University, Yen Nghia, Ha Dong, Hanoi 12116, Vietnam}
\date{\today}

\begin{abstract}
We study the constraint of the non-SUSY AdS conjecture on the three-dimensional vacua obtained from the compactification of the Standard Model coupled to Einstein gravity on a circle where the three-dimensional components of the four-dimensional metric are general functions of both non-compact and compact coordinates. From studying the wavefunction profile of the three-dimensional metric in the compactified dimension, we find that the radius of the compactified dimension must be quantized. Consequently, the three-dimensional vacua are constrained by not only the non-SUSY AdS conjecture but also the quantization rule of the circle radius, leading to both upper and lower bounds for the mass of the lightest neutrino as $\sqrt{2}\leq m_\nu/\sqrt{\Lambda_4}<\sqrt{3}$ where $\Lambda_4\simeq5.06\times10^{-84}$ GeV$^2$ is the observed cosmological constant. This means that the lightest neutrino should have a mass around $10^{-32}$ eV or it would be approximately massless. With this prediction, we reconstruct the light neutrino mass matrix that is fixed by the neutrino oscillation data and in terms of three new mixing angles and six new phases for both the normal ordering and inverted ordering. In the situation that the light neutrino mass matrix is Hermitian, we calculate its numerical value in the $3\sigma$ range.
\end{abstract}

\maketitle

\section{Introduction}

The observations of the neutrino oscillations have indicated the non-zero mass of the neutrinos but they only provide information about the difference of the squared neutrino masses given by \cite{Esteban2020}
\begin{eqnarray}
 \Delta m^2_{21}&=&7.39^{+0.21}_{-0.20}\times10^{-5} \text{eV}^2,\nonumber\\   
 \Delta m^2_{31}&=&2.525^{+0.033}_{-0.032}\times10^{-3} \text{eV}^2 \ \ \ \ \text{Normal ordering (ON)},\\
 \Delta m^2_{32}&=&-2.512^{+0.034}_{-0.032}\times10^{-3} \text{eV}^2 \ \ \ \ \text{Inverted ordering (IN)}.\nonumber
\end{eqnarray}
In this way, the present neutrino experiments cannot tell us the mass of the lightest neutrino by which the mass of three neutrino generations can be determined. The determination of the neutrino masses is very important to understand the role of neutrinos in the evolution of the universe and the formation of large structures (like galaxies) because the mass density of the universe would obtain significant contributions from the very light neutrinos \cite{Lesgourgues2012}.  

It is interesting in recent years that the consistency of the low-energy effective theories with quantum gravity, related to the swampland program \cite{Vafa2005,Ooguri2007,Palti2019}, can give constraints on the mass of the lightest neutrino. The Standard Model (SM) coupled to Einstein gravity is clearly a good effective theory in the low-energy regime and hence it must be consistent with quantum gravity. This means that the effective field theories that are derived from its dimensional reduction are too. On the other hand, the lower-dimensional vacua in the landscape of the SM coupled to Einstein gravity would be subject to the constraint of the swampland conjectures. In Refs. \cite{Martin-Lozano2017,Hamada2017}, by considering the compactification of the SM coupled to Einstein gravity on a circle \cite{Arkani-Hamed2007}, the constraint of the non-SUSY AdS conjecture (stating that stable non-SUSY AdS vacua are inconsistent with quantum gravity or belong to the swampland) imposes an upper bound on the mass of the lightest neutrino as $m_{\nu_{1(3)}}\lesssim(M^2_{\text{Pl}}\Lambda_4)^{1/4}\sim10^{-3}$ eV where $M_{\text{Pl}}\sim10^{19}$ GeV is the Planck energy scale and $\Lambda_4\simeq5.06\times10^{-84}$ GeV$^2$ is the observed cosmological constant \cite{Tanabashi2018}.

In this work, we will revisit the circle compactification of the SM coupled to Einstein gravity and the application of the non-SUSY AdS conjecture to the corresponding three-dimensional vacua where unlike the previous investigations in the literature we do not restrict the three-dimensional components of the four-dimensional metric to be the functions of the non-compact coordinates only. But, we consider that they are general functions of both non-compact and compact coordinates. By using the most general setting of the circle compactification described by a U(1) principal bundle, we expand the Einstein-Hilbert (EH) action plus a positive cosmological constant in terms of the three-dimensional metric, U(1) gauge vector, and radion fields, represented in Sec. \ref{CComp}. From studying the wavefunction profile of the three-dimensional metric in the compactified dimension, we show in Sec. \ref{Radquan} that the radius of the compactified dimension must be quantized due to its circle topology. In Sec. \ref{nonSUSY}, we indicate that the quantization rule of the size of the compactified dimension and the constraint of the non-SUSY AdS conjecture on the three-dimensional vacua predict a mass range for the lightest neutrino as $\sqrt{2}\leq m_\nu/\sqrt{\Lambda_4}<\sqrt{3}$, which means that the mass of the lightest neutrino should be around $10^{-32}$ eV or it would be well-approximately massless. From this prediction, in Sec. \ref{LNMM} we reconstruct the light neutrino mass matrix which can be given by the numerical values in the $3\sigma$ range if the light neutrino mass matrix is Hermitian.

\section{\label{CComp} Circle compactification}
We start introducing the four-dimensional action describing the SM coupled to Einstein gravity as follows 
\begin{eqnarray}
S&=&\frac{M^2_{\text{Pl}}}{2}\int d^4x\sqrt{-g_4}\left[\mathcal{R}^{(4)}-\Lambda_4\right]+S_{\text{SM}},\label{SMEHLa-act}
\end{eqnarray}
where $\mathcal{R}^{(4)}$ is the Ricci scalar and $S_{\text{SM}}$ denotes the action of the SM. As we will see later, only the lightest field of the SM which is the lightest neutrino plays a role under the constraint of the non-SUSY AdS conjecture because it significantly contributes to the radion potential whose minimum would determine a landscape of the three-dimensional vacua. In this sense, the relevant action of the SM is given by
\begin{eqnarray}
S_{\text{SM}}=\int d^4x\sqrt{-g_4}\bar{\nu}_{1(3)}\left(i\gamma_\alpha{e_\alpha}^\mu D_\mu-m_{\nu_{1(3)}}\right)\nu_{1(3)}+\cdots, 
\end{eqnarray}
where ${e_\alpha}^\mu$ and $D_\mu$ denote the vierbein and the covariant derivative, respectively, and the ellipsis refers to the rest of the SM fields which can be ignored under the constraint of the non-SUSY AdS conjecture.

Because our starting point is a vacuum configuration of positive energy in four-dimensional Einstein gravity coupled to the SM, we discuss the constraints on dS spacetime background coming from other swampland conjectures. Motivated by the difficulties in constructing dS vacuum solutions in string theory, it has been conjectured that string theory does not admit dS vacua, or in other words, dS vacua would be in the swampland. The idea that no dS vacua are consistent with quantum gravity is quantitatively formulated by the (refined) dS conjecture \cite{Obied2018,Ooguri2019,Krishnan075} that requires the scalar potential of the effective field theories coupled with Einstein gravity to satisfy the following bound
\begin{eqnarray}
|\nabla V|\geq\frac{c}{M_{\text{Pl}}}V \ \ \ \ \text{or} \ \ \ \ \text{min}\left(\nabla_i\nabla_j V\right)\leq -\frac{c'}{M^2_{\text{Pl}}}V,\label{dsconje}
\end{eqnarray}
where $c$ and $c'$ are the $\mathcal{O}(1)$ positive constants. It is obvious that the dS conjecture forbids (meta-)stable dS vacua because they violate both of the inequalities given in (\ref{dsconje}). However, the dS conjecture is in direct tension with both inflation and $\Lambda$CDM model which are well consistent with the observational data. On the other hand, the possibility that dS vacua are inconsistent with quantum gravity is not ruled out by the observations. Therefore, the trans-Planckian censorship conjecture (TCC) was recently proposed as a swampland criterion to further relax the constraint of the dS conjecture \cite{Bedroya2020,Brandenberger2020,Brandenberger2021}. This conjecture forbids stable dS vacua but allows the existence of metastable dS vacua as long as their lifetime cannot last longer than $H^{-1}\log\left(M_{\text{Pl}}/H\right)$. With the lifetime in order of Hubble time, the universe may exist in a metastable dS vacuum for a long enough time \cite{Seo2020,Cai2021}. Hence, it can be approximated as a stable vacuum configuration where we can impose some swampland conjectures to set the constraints on the low-energy dynamics in the IR regime.

Now we consider the compactification of the SM coupled to Einstein gravity on a circle $S^1$ with $x^3$ and $x^3+2\pi$ identified where the setting of this compactification is described in the most general way by a principal bundle with the typical fiber U(1) \cite{Coquereaux1988,Bailin1987,Overduin1997,Nam2019,Nam2021}. Using this setting, we can write the metric endowed on the spacetime compactified on the circle $S^1$ as follows
\begin{eqnarray}
ds^2=g_{ij}dx^i dx^j+R^2\left[dx^3+\kappa A_i dx^i\right]^2.\label{KKmetric}
\end{eqnarray}
Here $g_{ij}$ is the three-dimensional metric, $A_i$ is the connection or the gauge vector field on the U(1) principal bundle and transforms under the general coordinate transformations as $A_i\rightarrow A_i-\partial_i\epsilon(x^j)/\kappa$, $R$ is the radion (or dilaton) field whose vacuum expectation value would fix the radius of the $S^1$ fiber, $i,j=0,1,2$ are the indices of the non-compact coordinates, and $\kappa$ is the gauge coupling. Note that, the three-dimensional field $g_{ij}$ and the radion field $R$ are in general dependent on both non-compact and compact coordinates. 

Replacing the ansatz (\ref{KKmetric}) into the Ricci scalar $\mathcal{R}^{(4)}$, the EH action plus the cosmological constant gives \cite{Nam2023-2}
\begin{eqnarray}
S_{\text{EH}}&=&\frac{M^2_{\text{Pl}}}{2}\int d^4x\sqrt{-g_4}\left[\hat{\mathcal{R}}+\frac{1}{4R^2}\left(\partial_3g^{ij}\partial_3 g_{ij}+g^{ij}g^{kl}\partial_3 g_{ij}\partial_3 g_{kl}\right)-\frac{\kappa^2R^2}{4}F_{ij}F^{ij}-\Lambda_4\right],\label{Gravact} 
\end{eqnarray}
where $\hat{\mathcal{R}}$ and the field strength tensor $F_{ij}$ of the U(1) gauge field are given by
\begin{eqnarray}
\hat{\mathcal{R}}&\equiv&g^{ij}\left(\hat{\partial}_k\hat{\Gamma}^k_{ji}-\hat{\partial}_j\hat{\Gamma}^k_{ki}+\hat{\Gamma}^k_{ji}\hat{\Gamma}^l_{lk}-\hat{\Gamma}^k_{li}\hat{\Gamma}^l_{jk}\right),\\
\hat{\Gamma}^k_{ij}&\equiv&\frac{g^{kl}}{2}\left(\hat{\partial}_i g_{jl}+\hat{\partial}_j g_{il}-\hat{\partial}_l g_{ij}\right),\\
F_{ij}&=&\partial_i A_j-\partial_j A_i
\end{eqnarray}
with $\hat{\partial}_i\equiv\partial_i-\kappa A_i\partial_3$. A novel point that appears in the gravity action (\ref{Gravact}) upon the circle compactification is the presence of the second term. This is due to the fact that in the calculation we have not restricted the three-dimensional field $g_{ij}$ to be the functions of $x^i$ only as considered in the previous investigations about the circle compactification of the SM coupled to Einstein gravity \cite{Martin-Lozano2017,Hamada2017,Arkani-Hamed2007,Arnold2010,Fornal2011,Valenzuela2017,Gonzalo2018a,Gonzalo2018b}. As we will indicate late, this novel term would lead to a significant change to the constraint of the lightest neutrino mass. In addition, it should be noted that no kinetic term related to the first-order derivatives of the radion field appears in the expansion of the Ricci scalar $\mathcal{R}^{(4)}$ because the $S^1$ fiber has a zero curvature. However, the kinetic term of the radion field will appear in the Einstein frame obtained by rescaling the three-dimensional metric.

\section{\label{Radquan} Quantization of the circle radius}

\subsection{Equations of motion for the 3D components of the bulk metric}
In this subsection, we will write explicitly a full set of equations of motions for the three-dimensional components of the bulk metric in order to show that our setup is actually self-consistent. We would like to note that in Ref. \cite{Nam2023-2} the components $g_{ij}$, $A_i$, and $R$ are considered to be the general functions of the non-compact coordinates as well as the fourth compact circle coordinate. However, in the calculation of the scalar curvature $\mathcal{R}^{(4)}$ as given in Appendix A in Ref. \cite{Nam2023-2}, the terms relating to $\partial_3A_i$ automatically cancel together. As a result, they do not contribute to $\mathcal{R}^{(4)}$ because the contribution to the spacetime curvature caused by the topological non-triviality of the $U(1)$ principal bundle is measured by the curvature two-form $F_{ij}$. Therefore, in this work, we have not considered the dependence of the fourth compact circle coordinate entering through the U(1) field $A_i$. However, the equations of motion for the three-dimensional components of the bulk metric are basically similar to those derived in Ref. \cite{Nam2023-2} (see Appendix B of this reference).

The equations of motion for the three-dimensional metric $g_{ij}$ can be found via the variation $\delta_{g_{ij}}S_{\text{EH}}=0$ with $S_{\text{EH}}$ as given in Eq. (\ref{Gravact}) and they are given by
\begin{eqnarray}
\mathcal{R}-6\Lambda_4-\frac{1}{4R^2}\left[8g^{ij}\partial^2_3g_{ij}+5\partial_3 g^{ij}\partial_3g_{ij}+9\left(g^{ij}\partial_3 g_{ij}\right)^2\right]-\frac{\kappa^2R^2}{4}F_{kl}F^{kl}+f_1\left(\hat{\partial}_iR,\partial_3R\right)&=&0,\label{gij-eqs}\nonumber\\
\end{eqnarray}
where $\mathcal{R}$ and $\Gamma^k_{ij}$ are defined as follows
\begin{eqnarray}
\mathcal{R}&\equiv&g^{ij}\left(\partial_k\Gamma^k_{ji}-\partial_j\Gamma^k_{ki}+\Gamma^l_{ji}\Gamma^k_{kl}-\Gamma^l_{ki}\Gamma^k_{jl}\right),\\
\Gamma^k_{ij}&\equiv&\frac{g^{kl}}{2}\left(\partial_i g_{lj}+\partial_j g_{ki}-\partial_l g_{ij}\right),
\end{eqnarray}
and $f_1(\hat{\partial}_iR,\partial_3R)$ is a functional of $\hat{\partial}_iR$ and $\partial_3R$ and it vanishes when $R$ is a constant. From the variation $\delta_{A}S_{\text{EH}}=0$, we can find the equations of motion for the three-dimensional gauge field $A_i$ as follows
\begin{equation}
\nabla_i F^{ij}=-\frac{3\hat{\partial}_iR}{R}F^{ij}+f_2\left(\hat{\partial}_ig_{kl},\partial_3g_{kl}\right)A^j,\label{Ai-equ}
\end{equation}
where $f_2(\hat{\partial}_ig_{kl},\partial_3g_{kl})$ is a functional of $\hat{\partial}_ig_{kl}$ and $\partial_3g_{kl}$. With respect to the radion field, as seen in the next section, it would get a three-dimensional effective potential which is generated by the cosmological constant term and the Casimir energy coming from loops wrapping the fourth compactified dimension. Hence, we obtain the equation of motion for the radion field $R$ from the variation of the three-dimensional action given in Eq. (\ref{4Deffact}) in terms of $R$. It is given by
\begin{eqnarray}
\square R&=&\frac{1}{R}\left(\partial_iR\right)^2+\frac{R^2}{2M_3}\frac{\partial V(R)}{\partial R}+f_3(A_i,F_{ij}),\label{rad-equ}
\end{eqnarray}
where $V(R)$ is the three-dimensional effective potential of the radion field and $f_3(A_i,F_{ij})$ is a functional of $A_i$ and $F_{ij}$. Note that, the functional $f_3(A_i,F_{ij})$ is zero with the vanishing of the curvature two-form $F_{ij}$. 

The three-dimensional effective potential of the radion field $R$ allows us to stabilize the size of the fourth compactified dimension. This implies that we consider the theory in the vacuum $R=\text{constant}$ which corresponds to the minimum of the radion potential. From Eq. (\ref{rad-equ}), the solution with $R=\text{constant}$ is only satisfied when $f_3(A_i,F_{ij})=0$ which leads to $F_{ij}=0$ corresponding to $A_i=0$ in a proper gauge. We can easily check that the solution $R=\text{constant}$ and $A_i=0$ also is automatically satisfied by the equations of motion for the three-dimensional gauge field $A_i$ as given in Eq. (\ref{Ai-equ}). Therefore, the KK gauge field $A_i$ should be vanishing in the background where the radion field is constant which is physically fixed by the minimum of the radion potential.

\subsection{Quantization rule}
In the vacuum $R=\text{constant}$ and $A_i=0$, we find the equations of motion for the three-dimensional metric $g_{ij}$ as follows
\begin{eqnarray}
\mathcal{R}-6\Lambda_4-\frac{1}{4R^2}\left[8g^{ij}\partial^2_3g_{ij}+5\partial_3 g^{ij}\partial_3g_{ij}+9\left(g^{ij}\partial_3 g_{ij}\right)^2\right]&=&0,\label{3Dtensor-equ}
\end{eqnarray}
It is interesting that Eq. (\ref{3Dtensor-equ}) can be solved by separating the variables as follows
\begin{eqnarray}
g_{ij}(x^i,x^3)=\chi(x^3)g^{(3)}_{ij}(x^i),\label{varsepa}
\end{eqnarray}
where $g^{(3)}_{ij}(x^i)$ is identified as the metric defining the line element of the effective three-dimensional spacetime and $\chi(x^3)$ describes its wavefunction profile in the compactified dimension. We substitute Eq. (\ref{varsepa}) into Eq. (\ref{3Dtensor-equ}), then we derive the following equations
\begin{eqnarray}
\mathcal{R}^{(3)}&=&6\Lambda_3,\label{effEinsEq}\\
\chi''+\frac{11}{4}\frac{\chi'^2}{\chi}+\frac{\Lambda_4}{R^{-2}}\chi&=&\frac{\Lambda_3}{R^{-2}},\label{chiEq}
\end{eqnarray}
where $\mathcal{R}^{(3)}$ is the Ricci scalar of the effective three-dimensional spacetime whose geometry is determined by Eq. (\ref{effEinsEq}). The source of the geometry of the effective three-dimensional spacetime is a cosmological constant $\Lambda_3$ originating from the dynamics of the three-dimensional metric in the compactified dimension. By solving Eq. (\ref{chiEq}), one can find the wavefunction profile of the three-dimensional metric in the compactified dimension. It is important to note here that the equation for the wavefunction profile is nonlinear due to the nonlinear nature of the three-dimensional metric. This means that the solution for the three-dimensional metric should not be expressed as a linear combination of the partial solutions.

Due to the presence of the nonlinear term (which is the second term on the left-hand side) in Eq. (\ref{chiEq}), it is not easy to find an analytical solution for the wavefunction profile of the three-dimensional metric. But, in the case of $\Lambda_3/R^{-2}\ll1$ [which can be realized from Eq. (\ref{radquan}) and Table. \ref{tab-n-La3-mnu}], Eq. (\ref{chiEq}) can be perturbatively solved in the order of $\Lambda_3/R^{-2}$. At the leading order, we can find an analytical solution for the wavefunction profile of the three-dimensional metric as follows
\begin{eqnarray}
\chi(x^3)=\left[\frac{1}{2}\left\{1+\cos\left(R\sqrt{15\Lambda_4}x^3\right)\right\}\right]^{2/15}.
\end{eqnarray}
Importantly, the circle topology requires that the wavefunction profile of the three-dimensional metric must be periodic with the period of $2\pi$ as $\chi(x^3)=\chi(x^3+2\pi)$. This leads to a quantization condition for the circle radius and cosmological constant as follows
\begin{eqnarray}
R=\sqrt{\frac{1}{15\Lambda_4}}n.\label{radquan} \ \  \text{with}\ \  n=1,2,3,\cdots.
\end{eqnarray}

In this way, the nontrivial behavior of the wavefunction profile of the three-dimensional metric in the compactified dimension requires both the radius of the compactified dimension and the four-dimensional cosmological constant to be quantized. In other words, their value is not arbitrary but only obtains the discrete values satisfying the quantization rule (\ref{radquan}). Such a quantization has been used to interpret the radiative stability of tiny observed cosmological constant under the quantum corrections \cite{{Nam2023-2}} as well as to construct microscopic configurations for observed black holes \cite{Nam2023-4}

\section{\label{nonSUSY} Constraint of non-SUSY AdS conjecture}

The compactification of the SM coupled to GR on the circle $S^1$ would lead to a landscape of vacua which corresponds to the extrema of the radion potential. Here, the radion potential is generated by the cosmological constant and the one-loop correction of the light particles. Depending on the three-dimensional cosmological constant $\Lambda_3$ and the mass of the light particles, this landscape of vacua includes both the dS and AdS geometries which would be an object for the constraint of the non-SUSY AdS conjecture.

The three-dimensional effective action of Einstein gravity derived upon the circle compactification in the Einstein frame is given by
\begin{eqnarray}
S_{\text{3D}}&\supset&\int d^3x\sqrt{-g_3}\left[\frac{M_3}{2}\mathcal{R}^{(3)}-2M_3\left(\frac{\partial_\mu R}{R}\right)^2-M_3\left(\frac{r}{R}\right)^2\Lambda_3\right],\label{4Deffact}
\end{eqnarray}
where we have rescaled the three-dimensional metric as $g^{(3)}_{ij}\rightarrow\Omega^{-2}g^{(3)}_{ij}$ with $\Omega=R/r$ (the parameter $r$ is introduced to keep the rescaled three-dimensional metric dimensionless and it would be fixed equal to the vacuum expectation value of the radion field), $g_3\equiv\text{det}\left[g^{(3)}_{ij}\right]$, and the three-dimensional Planck energy scale is identified as
\begin{eqnarray}
M_3&\equiv&rM^2_{\text{Pl}}\int^\pi_{-\pi}dx^3\chi^{1/2}\nonumber\\
&=&\frac{2\sqrt{\pi}\Gamma(17/30)}{\Gamma(16/15)}rM^2_{\text{Pl}}.
\end{eqnarray}
The last term in the action (\ref{4Deffact}) is the tree-level potential of the radion field generated by the dynamics of the three-dimensional metric along the compactified dimension. In addition, the one-loop quantum corrections would contribute to the radion potential as 
\begin{eqnarray}
V_{\text{1L}}(R)&=&\sum_i(-1)^{s_i}n_iR\left(\frac{r}{R}\right)^3\rho_i(R)\int^\pi_{-\pi}dx^3\chi^{\frac{3}{2}}.\ \
\end{eqnarray}
Here $s_i$ is equal to $0(1)$ for the fermions(bosons), $n_i$ is the number of degrees of freedom corresponding to the $i$-th particle, and the Casimir energy density with respect to the $i$-th particle is given by \cite{Arkani-Hamed2007}
\begin{eqnarray}
\rho_i(R)=\sum^{\infty}_{n=1}\frac{2m^4_i}{(2\pi)^2}\frac{K_2(2\pi nm_iR)}{(2\pi nm_iR)^2},
\end{eqnarray}
where $m_i$ and $K_2(z)$ are the mass of the $i$-th particle and the modified Bessel function, respectively. It should be noted here that due to the function $K_2(z)$ suppressed for $z\ll1$ the particles with their mass which is much larger than $R^{-1}$ do not contribute significantly to the one-loop term of the radion potential and hence we can ignore their contribution. On the other hand, only the light degrees of freedom contribute significantly to $V_{\text{1L}}(R)$.

From the tree and loop level contributions derived above, we can write the radion potential $V(R)$ by expanding it in terms of $m_iR$ for $m_iR\ll1$ as follows
\begin{eqnarray}
\frac{V(R)}{2\sqrt{\pi}r^3}&\simeq&\left[\frac{\Gamma(17/30)}{\Gamma(16/15)}\frac{M^2_P\Lambda_3}{R^2}+\frac{1}{16\pi^2}\frac{\Gamma(7/10)}{\Gamma(6/5)}\frac{1}{R^6}\sum_i(-1)^{s_i}n_i\left\{\frac{1}{90}-\frac{(m_iR)^2}{6}+\frac{(m_iR)^4}{48}\right\}\right].
\end{eqnarray}
By studying the behavior of the radion potential $V(R)$ when changing the 3D cosmological constant $\Lambda_3$, we find that $V(R)$ can exhibit a stable non-SUSY AdS vacuum for $\Lambda_3\leq0$ \cite{Gonzalo2021}. However, as indicated by the non-SUSY AdS instability conjecture, the stable non-SUSY AdS vacua should belong to the swampland, which means that
the SM coupled to Einstein gravity is inconsistent with the UV embedding in quantum gravity.
On the other hand, in order for the SM coupled to Einstein gravity coming from quantum gravity, the radion potential must develop a non-AdS 3D vacuum. This requires the following two conditions: (i) $\Lambda_3>0$ and (ii) the number of light fermionic degrees of freedom 
must be larger than the number of the massless bosonic degrees of freedom which are two from graviton and two from the photon.\footnote{The non-SUSY AdS conjecture is applied with an assumption that the UV non-perturbative instabilities that are transferred to the effective 3D theory are absent.} The experimental value $\Lambda_4\simeq5.06\times10^{-84}$ GeV$^2$ \cite{Tanabashi2018} and Eq. (\ref{radquan}) lead to $R\gtrsim1.15\times10^{41}$ GeV$^{-1}$ which is the mass order for the fermionic degrees of freedom contributing significantly to the radion potential. This means that only the lightest neutrino can contribute to the radion potential if it is light enough. But, its possible maximum degrees of freedom are four if it is a Dirac particle. As a result, the lightest neutrino itself is impossible to generate a non-AdS 3D vacuum. Therefore, there are additional light fermionic degrees of freedom that are beyond the SM with the masses in the order of $\sqrt{\Lambda_4}$. With such a tiny upper bound for the mass, we can consider an additional massless Dirac fermion. 

\begin{figure}[h]
\centering
\begin{tabular}{cc}
\includegraphics[width=0.5 \textwidth]{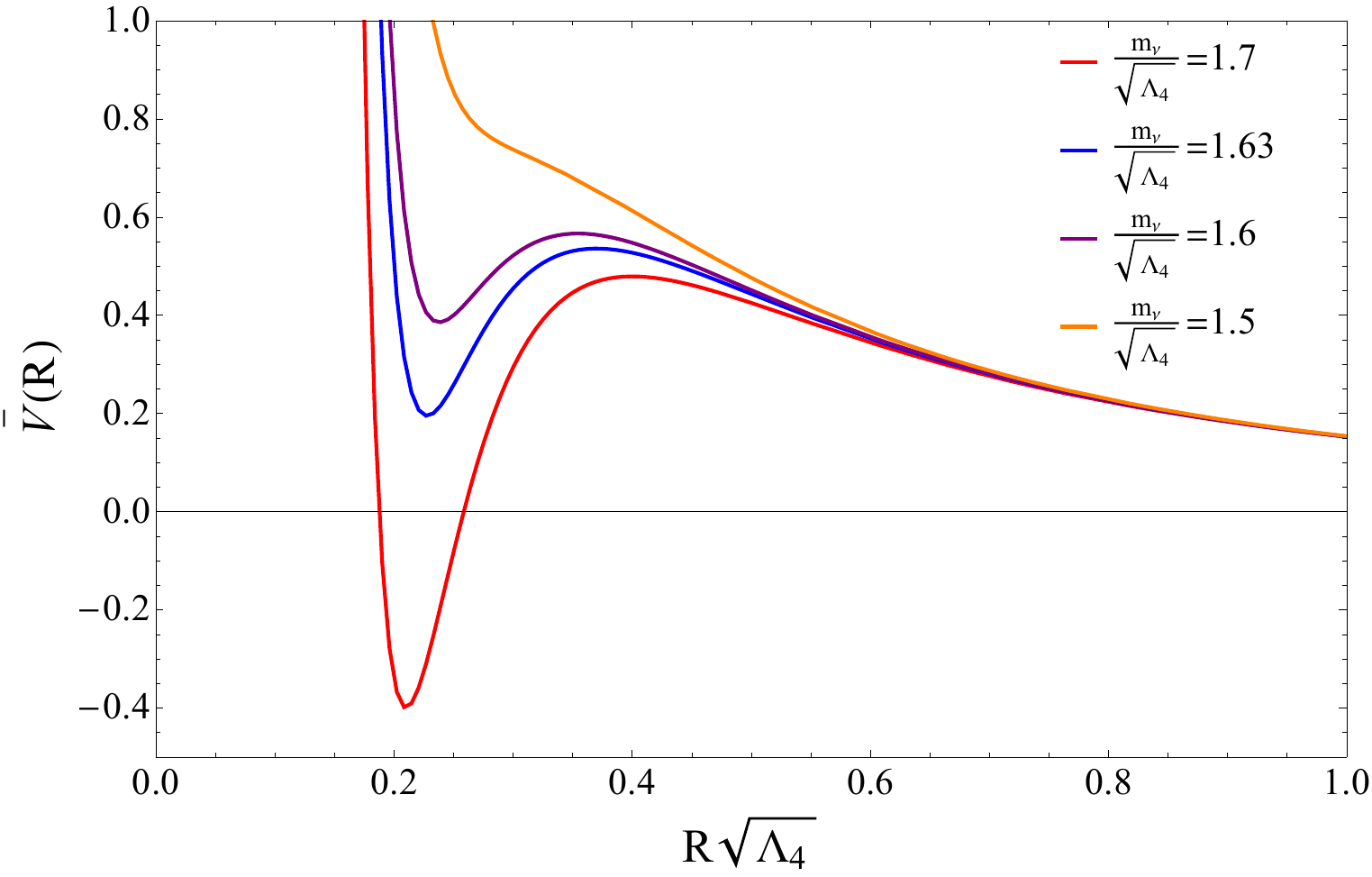}
\end{tabular}
 \caption{The behavior of the radion potential rescaled as $\bar{V}(R)\equiv V(R)/(2\sqrt{\pi}r^3\Lambda^6_4)$ in terms of the radius of one compactified spatial dimension rescaled as $R\sqrt{\Lambda_4}$ for various values of $m_\nu/\sqrt{\Lambda_4}$ with $M^2_{\text{Pl}}\Lambda_3/\Lambda^2_4=0.1$. The red and orange curves refer to the AdS vacuum and runaway dS vacuum, respectively. The blue and purple curves correspons to the dS vacuum.}\label{radpot}
\end{figure}

In Fig. \ref{radpot}, we depict the behavior of the radion potential in terms of the radius of the compactified dimension for various values of the lightest neutrino mass. 
In the region of the large radius, the contribution of the three-dimensional cosmological constant is dominant, positive, and decreases with the growth of $R$, hence the radion potential will go to zero from above when $R$ approaches infinitive. On the contrary, in the region of the small radius, the Casimir contribution of the bosons and fermions is dominant where the bosons and fermions contribute negatively and positively, respectively. However, due to the fermionic degrees of freedom larger than the bosonic one, the radion potential will approach infinity for $R\rightarrow0$. In particular, a minimum or runaway behavior (runaway dS vacua) is formed, depending on the mass of the lightest neutrino.

In the previous works \cite{Martin-Lozano2017,Hamada2017}, there is no constraint on the runaway behavior of the radion potential. On the other hand, no lower bound on the mass of the lightest neutrino is imposed, only the upper bound is required to guarantee the absence of the stable non-SUSY AdS vacuum. Interestingly, in the present work, the runaway behavior of the radion potential must be excluded by the quantization of the circle compactification. Indeed, in the large field region in which the gradients of the radion potential are very small, or in other words the radion field exhibits ultra-slow-rolling behavior, we can study the theory with the radius of the compactified dimension kept fixed. In this situation, as found above the quantization condition (\ref{radquan}) requires that the radius of the compactified dimension only obtains the discrete values. However, for the runaway behavior, the spectrum of the radius of the compactified dimension is continuous and hence it would violate the quantization condition (\ref{radquan}). This would allow us to impose a lower on the mass of the lightest neutrino as given by Eq. (\ref{lob-neumass}).

A minimum will develop in the situation in which the mass of the lightest neutrino is above the lower bound. From the minimum condition $\partial V(R)/\partial R=0$, we derive an expression for the value of $R$ corresponding to the minimum of the radion potential as follows
\begin{eqnarray}
R_{\text{min}}=\frac{2}{\left[20m^2_\nu+\sqrt{30(13m^4_\nu-64\pi^2\alpha M^2_{\text{Pl}}\Lambda_3)}\right]^{1/2}},\label{radius-min}
\end{eqnarray}
where $m_\nu$ refers to the mass of the lightest neutrino and
\begin{eqnarray}
\alpha\equiv\frac{\Gamma(17/30)}{\Gamma(16/15)}\frac{\Gamma(6/5)}{\Gamma(7/10)}.
\end{eqnarray}
 It is clear from Eq. (\ref{radius-min}) that a condition for the existence of the (meta)stable minimum of the radion potential is
 \begin{eqnarray}
 m^4_\nu>\frac{64}{13}\pi^2\alpha M^2_{\text{Pl}}\Lambda_3.\label{lob-neumass}
 \end{eqnarray}
As we argued above, the radius of the compactified dimension must be quantized according to Eq. (\ref{radquan}). Therefore, in order for the theory to be self-consistent, the value of $R$ at the minimum of the radion potential corresponding to the vacuum expectation value (vev) of the radion field must be equal to $n/\sqrt{15\Lambda_4}$. This would imply the constraint on the light particle spectrum of the effective field theories: with respect to the light particles that contribute to the effective potential of the radion field, their mass should not be arbitrary but only obtain allowed values so that the radius of the fourth compactified dimension obeys the quantization rule (16). More specifically in the context of the present work, by requiring $R_{\text{min}}=n/\sqrt{15\Lambda_4}$, the mass of the lightest neutrino should be quantized according to the rule determined by the following equation
\begin{eqnarray}
m_\nu=\frac{\sqrt{2}}{n}\left[60\Lambda_4-\sqrt{6\left(585\Lambda^2_4-8n^4\pi^2\alpha M^2_{\text{Pl}}\Lambda_3\right)}\right]^{1/2},\label{ligneumass-exp}
\end{eqnarray}
with the following condition
\begin{eqnarray}
\frac{n^4\pi^2\alpha M^2_{\text{Pl}}\Lambda_3}{\Lambda^2_4}<\frac{117}{64}.
\end{eqnarray}
Eq. (\ref{ligneumass-exp}) should be one of the essential points that are used to predict the mass of the light neutrino.

In addition, in order for a stable non-SUSY AdS vacuum that can not develop in the lower dimensions, the following condition must satisfy
\begin{eqnarray}
m^4_\nu\leq\frac{192}{29}\pi^2\alpha M^2_{\text{Pl}}\Lambda_3.\label{non-SUSY-AdS}
\end{eqnarray}
This constraint leads to a lower bound on the three-dimensional cosmological constant $\Lambda_3$ as follows
\begin{eqnarray}
\frac{n^4\pi^2\alpha M^2_{\text{Pl}}\Lambda_3}{\Lambda^2_4}\geq\frac{29}{48}.
\end{eqnarray}

\begin{figure}[h]
 \centering
\begin{tabular}{cc}
\includegraphics[width=0.5 \textwidth]{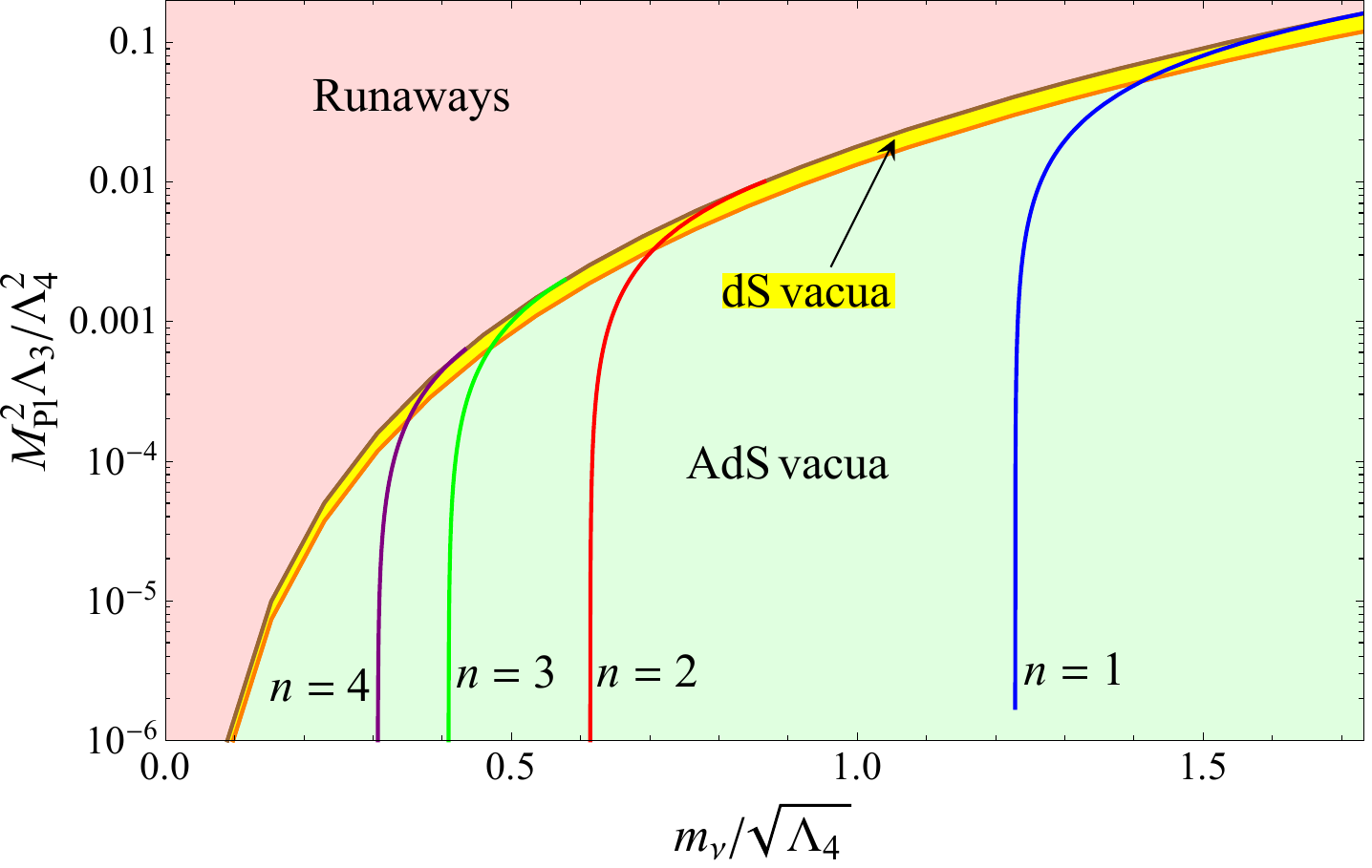}
\end{tabular}
 \caption{The lightest neutrino mass versus the three-dimensional cosmological constant under the quantization rule (\ref{radquan}) and the constraint of the non-SUSY AdS conjecture.}\label{La3-mnu}
\end{figure}

In summary, the quantization rule (\ref{radquan}) and the non-SUSY AdS conjecture impose both the upper and lower bounds on the three-dimensional cosmological constant $\Lambda_3$ and the mass of the lightest neutrino as
\begin{eqnarray}
\frac{29}{48n^4\pi^2\alpha}&\leq&\frac{M^2_{\text{Pl}}\Lambda_3}{\Lambda^2_4}<\frac{117}{64n^4\pi^2\alpha},\\
\frac{\sqrt{2}}{n}&\leq&\frac{m_\nu}{\sqrt{\Lambda_4}}<\frac{\sqrt{3}}{n}.\label{lnmass-bound}
\end{eqnarray}
For more explicitly, we show the allowed values of the three-dimensional cosmological constant versus those of the lightest neutrino mass (which are scaled by the 4D cosmological constant) for some values of the quantum number $n$ in Fig. \ref{La3-mnu} and in Table. \ref{tab-n-La3-mnu}. 
Their allowed values correspond to the curves (which are blue, red, green, and purple for $n=1$, $2$, $3$, and $4$, respectively, for example) belonging to the yellow region. 
\begin{table}[h]
\begin{tabular}{c|c|c} 
			\hline
			\hline
			\ \ $n$ \ \ & \ \ $m_\nu/\sqrt{\Lambda_4}$ \ \ & \ \ $M^2_{\text{Pl}}\Lambda_3/\Lambda^2_4$\ \  \\ \hline\hline
			1 &\ \ $(1.414,1.732)\ \ $ & $(0.053,0.161)$\\ \hline
			2 & $(0.707,0.866)$ & $(3.3\times10^{-3},0.01)$\\ \hline
			3 & $(0.471,0.577)$ & $\ \ (6.5\times10^{-4},2\times10^{-3})\ \ $\\ \hline
			4 & $(0.354,0.433)$ & $(2\times10^{-4},6\times10^{-4})$\\ \hline
\end{tabular}
\caption{The predicted value range for the lightest neutrino mass and the three-dimensional cosmological constant for some values of the quantum number $n$.}\label{tab-n-La3-mnu}
\end{table}
We observe that the allowed region of the lightest neutrino mass and the 3D cosmological constant would narrow when the quantum number $n$ increases. The largest allowed region of the lightest neutrino mass which corresponds to $n=1$ is only $(1.414\sqrt{\Lambda_4},1.732\sqrt{\Lambda_4})$, as seen in Table. \ref{tab-n-La3-mnu}, which is around $\sqrt{\Lambda_4}\sim10^{-32}$ eV. With such a tiny mass scale, the lightest neutrino can be well-approximately considered to be massless. Hence, our present scenario provides an extremely predictive picture of the lightest neutrino mass. This is an essential point to distinguish our present scenario from the previous investigations \cite{Martin-Lozano2017,Valenzuela2017} which predicted $m_\nu\lesssim(M^2_{\text{Pl}}\Lambda_4)^{1/4}\sim10^{-3}$ eV.

The mass bound for the lightest neutrino given in Eq. (\ref{lnmass-bound}) is obtained from the expression of light neutrino mass given in Eq. (\ref{ligneumass-exp}) where the minimum of the radion potential is constrained by the non-SUSY AdS conjecture and the vacuum expectation value of the radion field is quantized by the condition (\ref{radquan}). This quantization condition is derived by solving the wavefunction profile $\chi(x^3)$ of the three-dimensional metric with the periodic condition, which relies essentially on the factorization (\ref{varsepa}). In this sense, it could be thought that the mass bound for the lightest neutrino is just the consequence of the factorization of the three-dimensional metric and the constraint of the non-SUSY AdS conjecture.

\section{\label{LNMM} Light neutrino mass matrix}
With the mass of the lightest neutrino predicted and the data of the neutrino oscillation \cite{Esteban2020}, we can determine the light neutrino mass matrix $M_\nu$ as
\begin{eqnarray}
M_\nu=U_L\text{diag}\left(m_{\nu_1},m_{\nu_2},m_{\nu_3}\right)U^\dagger_R,
\end{eqnarray}
where $m_{\nu_1}$, $m_{\nu_2}$, and $m_{\nu_3}$ are the masses of the light neutrinos given for the case of NO
\begin{eqnarray}
m_{\nu_1}\approx0, \ \ m_{\nu_2}=\sqrt{m^2_{\nu_1}+\Delta m^2_{21}},\ \ m_{\nu_3}=\sqrt{m^2_{\nu_1}+\Delta m^2_{31}}      
\end{eqnarray}
and for the case of IO
\begin{eqnarray}
m_{\nu_1}=\sqrt{m^2_{\nu_3}+\Delta m^2_{23}-\Delta m^2_{21}}, \ \ m_{\nu_2}=\sqrt{m^2_{\nu_3}+\Delta m^2_{23}},\ \  m_{\nu_3}\approx0,    
\end{eqnarray}
and $U_{L,R}$ are two unitary matrices used to diagonalize the light neutrino mass matrix which is a general $3\times3$ complex matrix. 

We are interested in the situation in which the charged lepton mass matrix is diagonal which appears in the models beyond the SM with the lepton generations carrying the different charges under new gauge symmetries, for instance, the $U(1)_{L_\mu-L_\tau}$ model \cite{Foot1991,XHe1991,Foot1994,Altmannshofer2016}. In this situation, the diagonalizing matrix $U_L$ is determined by the lepton mixing matrix $U_{\text{PMNS}}$ as follows
\begin{eqnarray}
U_L=U_{\text{PMNS}}=\left(%
\begin{array}{ccc}
  1 & 0 & 0 \\
  0 & c_{23} & s_{23} \\
  0 & -s_{23} & c_{23}\\
\end{array}%
\right)\left(%
\begin{array}{ccc}
  c_{13} & 0 & s_{13}e^{-i\delta} \\
  0 & 1 & s_{23} \\
  -s_{13}e^{i\delta} & 0 & c_{13}\\
\end{array}%
\right)\left(%
\begin{array}{ccc}
  c_{12} & s_{12} & 0 \\
  -s_{12} & c_{12} & 0 \\
  0 & 0 & 1 \\
\end{array}%
\right)P,
\end{eqnarray}
where $c_{ij}$ and $s_{ij}$ denote the cosine and sine of the mixing angles, respectively, $\delta$ refers to the CP phase of the lepton sector and $P$ is a diagonal matrix related to the Majorana phases. For the neutrinos of the Dirac nature, the Majorana phases are zero, and thus $P=\mathbb{1}$. The best-fit values of the mixing angles and the CP violation phase in the $3\sigma$ range are given in Table. \ref{3sig-mixang}
\begin{table}[h]
\begin{tabular}{c|c|c} 
			\hline
			\hline
                & \ \  NO  \ \ & \ \ IO \ \ \\ \hline\hline
			\ \ $s^2_{12}$ \ \ & \ \ $0.275\rightarrow0.350$ \ \ & \ \ $0.275\rightarrow0.350$\ \  \\ \hline
			$s^2_{23}$ & $0.418\rightarrow0.627$  & $0.423\rightarrow0.629$ \\ \hline
			$s^2_{13}$ & $\ \ 0.02045\rightarrow0.02439 \ \ $ &\ \ $0.02068\rightarrow0.02463$ \ \ \\ \hline
			$\delta$ & $125^\circ\rightarrow392^\circ$ & $196^\circ\rightarrow360^\circ$ \\ \hline
\end{tabular}
\caption{The $3\sigma$ range for the mixing angles and the CP phase of the neutrino oscillation \cite{Esteban2020}.}\label{3sig-mixang}
\end{table}

The diagonalizing matrix $U_R$ can be parametrized by three angles (denoted by $\bar{\theta}_{12}$, $\bar{\theta}_{23}$, and $\bar{\theta}_{13}$) and six phases (denoted by $\omega_1$, $\omega_2$, $\omega_3$, $\omega_4$, $\omega_5$, and $\delta_R$). Then, we can write $U_R$ as follows \cite{Senjanovic2016}
\begin{eqnarray}
U_R&=&\left(%
\begin{array}{ccc}
  e^{i\omega_1} & 0 & 0 \\
  0 & e^{i\omega_2} & 0 \\
  0 & 0 & e^{i\omega_3}\\
\end{array}%
\right)\left(%
\begin{array}{ccc}
  1 & 0 & 0 \\
  0 & \bar{c}_{23} & \bar{s}_{23} \\
  0 & -\bar{s}_{23} & \bar{c}_{23}\\
\end{array}%
\right)\left(%
\begin{array}{ccc}
  \bar{c}_{13} & 0 & \bar{s}_{13}e^{-i\delta_R} \\
  0 & 1 & s_{23} \\
  -\bar{s}_{13}e^{i\delta_R} & 0 & \bar{c}_{13}\\
\end{array}%
\right)\nonumber\\
&&\times\left(%
\begin{array}{ccc}
  \bar{c}_{12} & \bar{s}_{12} & 0 \\
  -\bar{s}_{12} & \bar{c}_{12} & 0 \\
  0 & 0 & 1 \\
\end{array}%
\right)\left(%
\begin{array}{ccc}
  e^{i\omega_4} & 0 & 0 \\
  0 & e^{i\omega_5} & 0 \\
  0 & 0 & 1\\
\end{array}%
\right),\label{UR-para}
\end{eqnarray}
where $\bar{c}_{ij}\equiv\cos\bar{\theta}_{ij}$ and $\bar{s}_{ij}\equiv\sin\bar{\theta}_{ij}$.

Using the lepton mixing matrix and the parametrization (\ref{UR-para}), the elements of the light neutrino mass matrix, denoted by $[M_\nu]_{ij}$, can be reconstructed by the neutrino oscillation data and in terms of three mixing angles and six phases in $U_R$. For the case of NO, these elements are given by the following analytical expressions
\begin{eqnarray}
 \left[M_\nu\right]_{11}&=&e^{-i\omega_1}\left[c_{13}\bar{c}_{13}s_{12}\bar{s}_{12}\sqrt{\Delta m^2_{21}}e^{-i\omega_5}+s_{13}\bar{s}_{13}\sqrt{\Delta m^2_{31}}e^{i(\delta_R-\delta)}\right],\nonumber\\
 \left[M_\nu\right]_{12}&=&e^{-i\omega_2}\left[c_{13}s_{12}\sqrt{\Delta m^2_{21}}e^{-i\omega_5}\left(\bar{c}_{12}\bar{c}_{23}-\bar{s}_{12}\bar{s}_{13}\bar{s}_{23}e^{-i\delta_R}\right)+\bar{c}_{13}s_{13}\bar{s}_{23}\sqrt{\Delta m^2_{31}}e^{-i\delta}\right],\nonumber\\
  \left[M_\nu\right]_{13}&=&e^{-i\omega_3}\left[c_{13}s_{12}\sqrt{\Delta m^2_{21}}e^{-i\omega_5}\left(\bar{c}_{12}\bar{s}_{23}+\bar{c}_{23}\bar{s}_{12}\bar{s}_{13}e^{-i\delta_R}\right)+\bar{c}_{13}\bar{c}_{23}s_{13}\sqrt{\Delta m^2_{31}}e^{-i\delta}\right],\nonumber\\
\left[M_\nu\right]_{21}&=&e^{-i\omega_1}\left[\bar{c}_{13}\bar{s}_{12}\sqrt{\Delta m^2_{21}}e^{-i\omega_5}\left(c_{12}c_{23}-s_{12}s_{13}s_{23}e^{i\delta}\right)+c_{13}s_{23}\bar{s}_{13}\sqrt{\Delta m^2_{31}}e^{i\delta_R}\right],\nonumber\\
\left[M_\nu\right]_{22}&=&e^{-i\omega_2}\left[\sqrt{\Delta m^2_{21}}e^{-i\omega_5}\left(c_{12}c_{23}-s_{12}s_{13}s_{23}e^{i\delta}\right)\left(\bar{c}_{12}\bar{c}_{23}-\bar{s}_{12}\bar{s}_{13}\bar{s}_{23}e^{-i\delta_R}\right)\right.\nonumber\\
&&\left.+c_{13}\bar{c}_{13}s_{23}\bar{s}_{23}\sqrt{\Delta m^2_{31}}\right],\\
\left[M_\nu\right]_{23}&=&e^{-i\omega_3}\left[-\sqrt{\Delta m^2_{21}}e^{-i\omega_5}\left(c_{12}c_{23}-s_{12}s_{13}s_{23}e^{i\delta}\right)\left(\bar{c}_{12}\bar{s}_{23}+\bar{c}_{23}\bar{s}_{12}\bar{s}_{13}e^{-i\delta_R}\right)\right.\nonumber\\
&&\left.+c_{13}\bar{c}_{13}\bar{c}_{23}s_{23}\sqrt{\Delta m^2_{31}}\right],\nonumber\\
\left[M_\nu\right]_{31}&=&e^{-i\omega_1}\left[c_{13}c_{23}\bar{s}_{13}\sqrt{\Delta m^2_{31}}e^{i\delta_R}-\bar{c}_{13}\bar{s}_{12}\sqrt{\Delta m^2_{21}}e^{-i\omega_5}\left(c_{12}s_{23}+c_{23}s_{12}s_{13}e^{i\delta}\right)\right],\nonumber\\
\left[M_\nu\right]_{32}&=&e^{-i\omega_2}\left[-\sqrt{\Delta m^2_{21}}e^{-i\omega_5}\left(c_{12}s_{23}+c_{23}s_{12}s_{13}e^{i\delta}\right)\left(\bar{c}_{12}\bar{c}_{23}-\bar{s}_{12}\bar{s}_{13}\bar{s}_{23}e^{-i\delta_R}\right)\right.\nonumber\\
&&\left.+c_{13}c_{23}\bar{c}_{13}\bar{s}_{23}\sqrt{\Delta m^2_{31}}\right],\nonumber\\
\left[M_\nu\right]_{33}&=&e^{-i\omega_3}\left[\sqrt{\Delta m^2_{21}}e^{-i\omega_5}\left(c_{12}s_{23}+c_{23}s_{12}s_{13}e^{i\delta}\right)\left(\bar{c}_{12}\bar{s}_{23}+\bar{c}_{23}\bar{s}_{12}\bar{s}_{13}e^{-i\delta_R}\right)\right.\nonumber\\
&&\left.+c_{13}c_{23}\bar{c}_{13}\bar{c}_{23}\sqrt{\Delta m^2_{31}}\right].\nonumber
\end{eqnarray}
We observe that the elements of the light neutrino mass matrix for the case of NO are independent of the phase $\omega_4$. While, for the case of IO, the elements of the light neutrino mass matrix reads
\begin{eqnarray}
\left[M_\nu\right]_{11}&=&c_{13}\bar{c}_{13}e^{-i\omega_1}\left(c_{12}\bar{c}_{12}\sqrt{\Delta m^2_{23}-\Delta m^2_{21}}e^{-i\omega_4}+s_{12}\bar{s}_{12}\sqrt{\Delta m^2_{23}}e^{-i\omega_5}\right),\nonumber\\
\left[M_\nu\right]_{12}&=&c_{13}e^{-i\omega_2}\left[s_{12}\sqrt{\Delta m^2_{23}}e^{-i\omega_5}\left(\bar{c}_{12}\bar{c}_{23}-\bar{s}_{12}\bar{s}_{13}\bar{s}_{23}e^{-i\delta_R}\right)-c_{12}\sqrt{\Delta m^2_{23}-\Delta m^2_{21}}e^{-i\omega_4}\right.\nonumber\\
&&\left.\times\left(\bar{c}_{23}\bar{s}_{12}+\bar{c}_{12}\bar{s}_{13}\bar{s}_{23}e^{-i\delta_R}\right)\right],\nonumber\\
\left[M_\nu\right]_{13}&=&c_{13}e^{-i\omega_3}\left[c_{12}\sqrt{\Delta m^2_{23}-\Delta m^2_{21}}e^{-i\omega_4}\left(\bar{s}_{12}\bar{s}_{23}-\bar{c}_{12}\bar{c}_{23}\bar{s}_{13}e^{-i\delta_R}\right)+s_{12}\sqrt{\Delta m^2_{23}}e^{-i\omega_5}\right.\nonumber\\
&&\left.\times\left(\bar{c}_{12}\bar{s}_{23}+\bar{c}_{23}\bar{s}_{12}\bar{s}_{13}e^{-i\delta_R}\right)\right],\nonumber\\
\left[M_\nu\right]_{21}&=&\bar{c}_{13}e^{-i\omega_1}\left[\bar{s}_{12}\sqrt{\Delta m^2_{23}}e^{-i\omega_5}\left(c_{12}c_{23}-s_{12}s_{13}s_{23}e^{i\delta}\right)-\bar{c}_{12}\sqrt{\Delta m^2_{23}-\Delta m^2_{21}}e^{-i\omega_4}\right.\nonumber\\
&&\left.\times\left(c_{23}s_{12}+c_{12}s_{13}s_{23}e^{i\delta}\right)\right],\nonumber\\
\left[M_\nu\right]_{22}&=&e^{-i\omega_2}\left[\sqrt{\Delta m^2_{23}-\Delta m^2_{21}}e^{-i\omega_4}\left(c_{23}s_{12}+c_{12}s_{13}s_{23}e^{i\delta}\right)\left(\bar{c}_{23}\bar{s}_{12}+\bar{c}_{12}\bar{s}_{13}\bar{s}_{23}e^{-i\delta_R} \right)\right.\nonumber\\
&&\left.+\sqrt{\Delta m^2_{23}}e^{-i\omega_5}\left(c_{12}c_{23}-s_{12}s_{13}s_{23}e^{i\delta}\right)\left(\bar{c}_{12}\bar{c}_{23}-\bar{s}_{12}\bar{s}_{13}\bar{s}_{23}e^{-i\delta_R}\right)\right],\\
\left[M_\nu\right]_{23}&=&-e^{-i\omega_3}\left[\sqrt{\Delta m^2_{23}-\Delta m^2_{21}}e^{-i\omega_4}\left(c_{23}s_{12}+c_{12}s_{13}s_{23}e^{i\delta}\right)\left(\bar{s}_{12}\bar{s}_{23}-\bar{c}_{12}\bar{c}_{23}\bar{s}_{13}e^{-i\delta_R}\right)\right.\nonumber\\
&&\left.+\sqrt{\Delta m^2_{23}}e^{-i\omega_5}\left(c_{12}c_{23}-s_{12}s_{13}s_{23}e^{i\delta}\right)\left(\bar{c}_{12}\bar{s}_{23}+\bar{c}_{23}\bar{s}_{12}\bar{s}_{13}e^{-i\delta_R}\right)\right],\nonumber\\
\left[M_\nu\right]_{31}&=&\bar{c}_{13}e^{-i\omega_1}\left[\bar{c}_{12}\sqrt{\Delta m^2_{23}-\Delta m^2_{21}}e^{-i\omega_4}\left(s_{12}s_{23}-c_{12}c_{23}s_{13}e^{i\delta}\right)-\bar{s}_{12}\sqrt{\Delta m^2_{23}}e^{-i\omega_5}\right.\nonumber\\
&&\left.\times\left(c_{12}s_{23}+c_{23}s_{12}s_{13}e^{i\delta}\right)\right],\nonumber\\
\left[M_\nu\right]_{32}&=&-e^{-i\omega_2}\left[\sqrt{\Delta m^2_{23}-\Delta m^2_{21}}e^{-i\omega_4}\left(s_{12}s_{23}-c_{12}c_{23}s_{13}e^{i\delta}\right)\left(\bar{c}_{23}\bar{s}_{12}+\bar{c}_{12}\bar{s}_{13}\bar{s}_{23}e^{-i\delta_R}\right)\right.\nonumber\\
&&\left.+\sqrt{\Delta m^2_{23}}e^{-i\omega_5}\left(c_{12}s_{23}+c_{23}s_{12}s_{13}e^{i\delta}\right)\left(\bar{c}_{12}\bar{c}_{23}-\bar{s}_{12}\bar{s}_{13}\bar{s}_{23}e^{-i\delta_R}\right)\right],\nonumber\\
\left[M_\nu\right]_{33}&=&e^{-i\omega_3}\left[\sqrt{\Delta m^2_{23}-\Delta m^2_{21}}e^{-i\omega_4}\left(s_{12}s_{23}-c_{12}c_{23}s_{13}e^{i\delta}\right)\left(\bar{s}_{12}\bar{s}_{23}-\bar{c}_{12}\bar{c}_{23}\bar{s}_{13}e^{-i\delta_R}\right)\right.\nonumber\\
&&\left.+\sqrt{\Delta m^2_{23}}e^{-i\omega_5}\left(c_{12}s_{23}+c_{23}s_{12}s_{13}e^{i\delta}\right)\left(\bar{c}_{12}\bar{s}_{23}+\bar{c}_{23}\bar{s}_{12}\bar{s}_{13}e^{-i\delta_R}\right)\right].\nonumber
\end{eqnarray}

It is interesting that it can take the mass matrix of the light neutrinos to be Hermitian when the generations of right-handed neutrinos transform universally. In Ref. \cite{LiuZhou2013}, the authors proved that a general complex mass matrix $M_\nu$ of the Dirac neutrinos can be decomposed as $M_\nu=S_\nu.V_\nu$ where $S_\nu$ and $V_\nu$ are a Hermitian matrix and a unitary matrix, respectively. And, by redefining the right-handed neutrino fields as $\nu'_R=V_\nu\nu_R$, the neutrino mass matrix would be Hermitian. In this situation, we have $U_R=U_L=U_{\text{PMNS}}$ which corresponds to $\bar{c}_{ij}=c_{ij}$ and $\bar{s}_{ij}=s_{ij}$, $\omega_1=\omega_2=\omega_3=\omega_4=\omega_5=0$, and $\delta_R=\delta$. As a result, the elements of the light neutrino mass matrix are fixed only by the neutrino oscillation data. The numerical value of the light neutrino mass matrix is given in the $3\sigma$ range for the case of NO
\begin{eqnarray}
\frac{M_\nu}{10^{-3}\text{eV}^2}= \left(%
\begin{array}{ccc}
  3.22\rightarrow4.32 & \left(0.82\rightarrow9.14\right)e^{i\left(-3.1\rightarrow1.9\right)} & \left(1.46\rightarrow9.2\right)e^{i\left(-3.13\rightarrow2.66\right)} \\
  (12)^* & 21.67\rightarrow36.14 & \left(16.41\rightarrow27.15\right)e^{i\left(-0.03\rightarrow0.03\right)} \\
  (13)^* & (23)^* & 19.71\rightarrow34.11\\
\end{array}%
\right),
\end{eqnarray}
and for the case of IO
\begin{eqnarray}
\frac{M_\nu}{10^{-3}\text{eV}^2}= \left(%
\begin{array}{ccc}
  43.79\rightarrow53.32 & \left(3.31\rightarrow7.54\right)e^{i\left(-2.62\rightarrow\pi\right)} & \left(3.05\rightarrow7.35\right)e^{i\left(-2.68\rightarrow\pi\right)} \\
  (12)^* & 17.12\rightarrow32.46 & \left(17.47\rightarrow32.28\right)e^{i\left(-\pi\rightarrow\pi\right)} \\
  (13)^* & (23)^* & 19.43\rightarrow35.24\\
\end{array}%
\right),
\end{eqnarray}
where $(ij)^*$ refers to the complex conjugate of the element $[M_\nu]_{ij}\times10^3$ eV$^{-2}$. Note that, the diagonal elements of the light neutrino mass matrix are real due to its Hermitian property.

\section{Conclusion}
The neutrino oscillation data only provides information about the difference of the squared neutrino masses without telling us the value of the lightest neutrino mass. Recently, it has been shown that the consistency of the compactification of the SM coupled to Einstein gravity to lower dimensions with quantum gravity can impose constraints on the mass of the lightest neutrino. The circle compactification may yield stable non-SUSY AdS vacua which are inconsistent with quantum gravity or belong to the swampland according to the non-SUSY AdS conjecture, depending on the light neutrino masses. Because the SM and Einstein gravity have described very well the observed world and hence they cannot be lying in the swampland, these dangerous AdS vacua must be absent if the mass of the lightest neutrino satisfies the following bound $m_\nu\lesssim\left(\rho_{4\text{D}}\right)^{1/4}$ where $\rho_{4\text{D}}\approx2.6\times10^{-47}$ GeV$^4$ is the observed vacuum energy density.

In the present work, we revisit the constraint of the non-SUSY AdS conjecture on the three-dimensional vacua which are obtained from the circle compactification of the SM coupled to Einstein gravity with the radion potential generated by the cosmological constant and the Casimir effect of the light particles. Unlike the previous studies where the three-dimensional components of the four-dimensional metric are restricted to be dependent on the non-compact coordinates only, we consider them to be the general functions of both non-compact and compact coordinates. From investigating the wavefunction profile of the three-dimensional metric in the compactified dimension, we find that the radius $R$ of the compactified dimension must be quantized by the following rule $R=n/\sqrt{15\Lambda_4}$ (due to the circle topology of the compactified dimension) where $n$ refers to the positive integers and $\Lambda_4=\rho_{4\text{D}}/M^2_{\text{Pl}}$ with $M_{\text{Pl}}$ being the observed Planck scale. Because of this quantization, the existence or the absence of dangerous three-dimensional AdS vacua is very sensitive to $\Lambda_4$ (instead of $\rho_{4\text{D}}$ as indicated in the previous works) and the light particles with the mass in the order of $\Lambda_4$. In addition, the quantization of the radius of the compactified dimension forbids the runaway behavior of the radion potential. Therefore, it is interesting that both the quantization rule and the non-SUSY AdS conjecture impose constraints on the three-dimensional vacua leading to an upper bound and a lower bound for the mass of the lightest neutrino as $\sqrt{2}\leq m_\nu/\sqrt{\Lambda_4}<\sqrt{3}$. This constraint is very predictive because the mass of the lightest neutrino is around $10^{-32}$ eV which implies that the lightest neutrino would be nearly massless.

With the well-approximately vanishing mass of the lightest neutrino, we reconstruct the light neutrino mass matrix in the situation that the charged lepton mass matrix is diagonal. In general, the light neutrino mass matrix is a $3\times3$ complex matrix with nine independent complex elements that can be diagonalized by the lepton mixing matrix (or the Pontecorvo-Maki-Nakagawa-Sakata matrix) and another unitary matrix. In this way, we fix the elements of the light neutrino mass matrix based on the neutrino oscillation data and in terms of three new mixing angles and six new phases for the cases of normal ordering (NO) and inverted ordering (IO). Interestingly, the mass matrix of the light neutrinos can be taken to be Hermitian when the generations of right-handed neutrinos transform universally by redefining the right-handed neutrino fields. In this case, we calculate the numerical value of the light neutrino mass matrix in the $3\sigma$ range for both NO and IO.

\end{document}